\documentclass{PoS}

\newcommand{\jpsi}{J/\psi}
\newcommand{\ups}{\Upsilon}
\newcommand{\cQ}{{\cal Q}}

\newcommand{\raa}{R_{\rm AA}}
\newcommand{\ie}{{\it i.e.}}
\newcommand{\eg}{{\it e.g.}}
\newcommand{\beq}{\begin{equation}}
\newcommand{\eeq}{\end{equation}}

\title{Heavy-Flavor Theory at ``Hard and Electromagnetic Probes 2018"}

\ShortTitle{Heavy Flavor}

\author{\speaker{Ralf Rapp}\thanks{A footnote may follow.}\\
        Cyclotron Institute and Dept. of Physics and Astronomy, Texas A\&M University, College Station, TX 77843-3366, USA\\
        E-mail: \email{rapp@comp.tamu.edu}}

\abstract{An overview is given of the theoretical developments on heavy quarks and quarkonia 
in heavy-ion collisions as reported at the ``Hard and EM Probes 2018" conference. Specifically, 
we address progress in the understanding of heavy-flavor diffusion and its hadronization, 
quarkonium transport and the extraction of quarkonium melting temperatures, energy loss of 
heavy quarks at high momentum, and their rescattering in small colliding systems.}

\FullConference{International Conference on Hard and Electromagnetic Probes of High-Energy Nuclear Collisions\\
		30 September - 5 October 2018\\
		Aix-Les-Bains, Savoie, France}

\begin{document}

\section{Introduction: Heavy Quarks and in-Medium QCD Force}
\label{sec_intro}
Heavy quarks are hard-produced probes of ultrarelativistic heavy-ion 
collisions (URHICs), yet they provide a unique access to the 
soft properties of the QCD medium. The large mass of heavy quarks, 
$m_Q\gg \Lambda_{\rm QCD}, T_{\rm c}$, essentially limits 
their production to the primordial collisions of nucleons, and implies 
theoretical simplifications in the 
microscopic description of their transport and spectral 
properties. The production systematics of heavy quarkonia 
(charmonia and bottomonia) have long been suggested as a means 
to study the medium modifications of the fundamental QCD force. 
The in-medium QCD force governs the temperature dependence of 
the quarkonium binding energies and dissociation rates. On the 
one hand, these properties control the transport of quarkonia 
through the fireball and thus their finally observed yields and momentum 
spectra in URHICs. On the other hand, the in-medium binding energies 
and dissociation rates (widths) are encoded in 
their spectral and correlation functions, and as such allow for 
quantitative tests against lattice-QCD (lQCD) data. The increasing 
variety of measured quarkonia in URHICs is now enabling systematic 
studies of ground and excited states that are starting to narrow down 
their transport parameters in phenomenological applications (see
Refs.~\cite{Andronic:2014sga,Rapp:2017chc,Ferreiro:2018umi} for recent 
reviews). 

A basic building block in the microscopic description of 
in-medium quarkonia, especially their dissociation widths, is 
the interaction of the constituent heavy anti-/quarks with the 
surrounding quark-gluon plasma (QGP). Since the in-medium 
quarkonium binding energies are typically of the order of a few 
hundred MeV or less, their dissociation processes are directly related 
to the coupling of low-momentum heavy quarks to the 
medium. In particular, the zero-momentum limit of the thermal 
relaxation rate of heavy quarks, $\gamma_Q$, determines their 
spatial diffusion coefficient, 
${\cal D}_s = T/(m_Q \gamma_Q(p=0))$, a fundamental transport 
parameter of QCD matter. This highlights the intimate connection 
between (chemical) quarkonium and (kinetic) heavy-quark (HQ) 
transport. A quantitative phenomenology of open heavy-flavor 
(HF) observables in URHICs requires, however, a good control over 
the 3-momentum dependence of the pertinent thermalization rates. 
A key issue here is the modelling of the medium constituents that 
the heavy quarks are interacting with -- after all, this at the 
core of the idea to utilize heavy quarks and quarkonia as a probe 
of the medium. Limiting cases are a weakly interacting gas of 
quasiparticles vs. a strongly interacting system without explicit 
quasiparticles within the AdS/CFT correspondence. Current phenomenological 
extractions~\cite{Rapp:2018qla,Gossiaux:2019mjc} of the HQ {\em scattering} 
rate, $\Gamma_Q$ (which for charm quarks is about an order of 
magnitude larger than the {\em thermalization} rate), yield values of 0.5\,GeV 
or more; this implies that the thermal medium partons, with effective masses 
of order $m_{q}\sim T\sim \Gamma_q$, 
are no longer well-defined quasiparticles! On the other hand, heavy 
quarks (in particular bottom), with $m_Q\gg T$, can remain good 
quasiparticles, which is a central reason why they are excellent 
``Brownian probes" of the strongly interacting QGP liquid. These 
arguments continue to apply to hadronization processes, which 
should emerge continuously in the transition from the QGP into 
the hadronic liquid. 
To take advantage of the close connection between HF transport 
and QGP structure, a unified treatment of the interactions in 
the heavy and light sector is desirable. 

The kinematics resulting from the HQ masses implies that the onset of 
radiative interactions is shifted to higher momenta, relative to light partons. 
For the latter the ``transition" from elastic to radiative interactions is 
quite possibly in a regime where they are near-thermalized in URHICs; this is 
presumably not the case for charm and bottom quarks; therefore, they retain a 
memory of their interaction history and thus can serve as a ``gauge" of the 
strength and type of their rescattering. Once radiation dominates, the quark 
mass dependence of energy loss can be studied.   
       
In the following, we discuss (in a non-exhaustive manner) the theoretical developments 
reported at ``Hard Probes '18" in the context of the above considerations, organized
into parts on HQ diffusion (Sec.~\ref{sec_hq}) and hadronization (Sec.~\ref{sec_hadro}),
quarkonium transport (Sec.~\ref{sec_onia}), high-$p_T$ energy 
loss (Sec.~\ref{sec_highpt}) and heavy flavor in small collision 
systems (Sec.~\ref{sec_small}), and conclude in 
Sec.~\ref{sec_concl}.

\section{Heavy-Quark Diffusion}
\label{sec_hq}
The propagation of heavy quarks through QCD matter is a multi-scale problem~\cite{Cao:2018};
while the hierarchies in kinematic variables can be reasonably well defined (\eg, for momentum 
transfer in the diffusive regime, $Q^2 \sim T^2 \sim K_T^2 = (p_T^2/2m_Q)^2 \ll p_T^2  \ll m_Q^2$),
the strong coupling of the QGP at moderate temperature renders the interactions intrinsically
nonperturbative. There is mounting evidence that
remnants of the confining force survive well above the pseudocritical temperature 
($T_{\rm pc}$); they could well be at the origin of the QGP's liquid-like properties at 
long wavelength, thereby providing a natural connection between the QGP's strong-coupling 
behavior and hadronization. 

The HF transport problem in URHICs can be decomposed into several  
components~\cite{Rapp:2018qla}: cold-nuclear-matter effects in HQ production, 
diffusion through a pre-equilibrium phase and the QGP, hadronization, hadronic
diffusion, and the ambient bulk evolution. 
An estimate of the effects of the pre-equilibrium phase, based on turbulent chromodynamic
fields, was presented in Ref.~\cite{Mrowczynski:2017kso}; it was 
found that the pertinent HQ transport coefficients are comparable to, or even larger than the 
equilibrium values at the same energy density. 
A special feature of these coefficients, rooted in the largely longitudinally oriented
chromo-electric fields, are rather pronounced rapidity 
dependences of the energy loss which may allow for experimental tests relative to the 
smoothly varying charged-particle densities figuring in the equilibrium description.  

Updates of an AdS/CFT-based HQ diffusion calculation in the QGP 
were reported in Ref.~\cite{Hambrock:2018sim}; the calculated 
$B$- and $D$-meson $\raa$'s, in comparison 
to LHC data, yield spatial diffusion coefficients in the range of ${\cal D}_s(2\pi T)$=1.2-1.7,
somewhat smaller than most other calculations (at least in part due to neglecting
recombination and the hadronic phase).  In addition, a constant momentum diffusion 
coefficient, $D_p$, was clearly favored over a $p$-dependent one, implying that the 
thermalization rate falls off as $\gamma_Q\sim 1/E$, quite similar to QCD-based 
models~\cite{Rapp:2018qla}. 

\begin{figure}[t]
\vspace{-0.5cm}
\begin{center}
\includegraphics[width=0.7\textwidth]{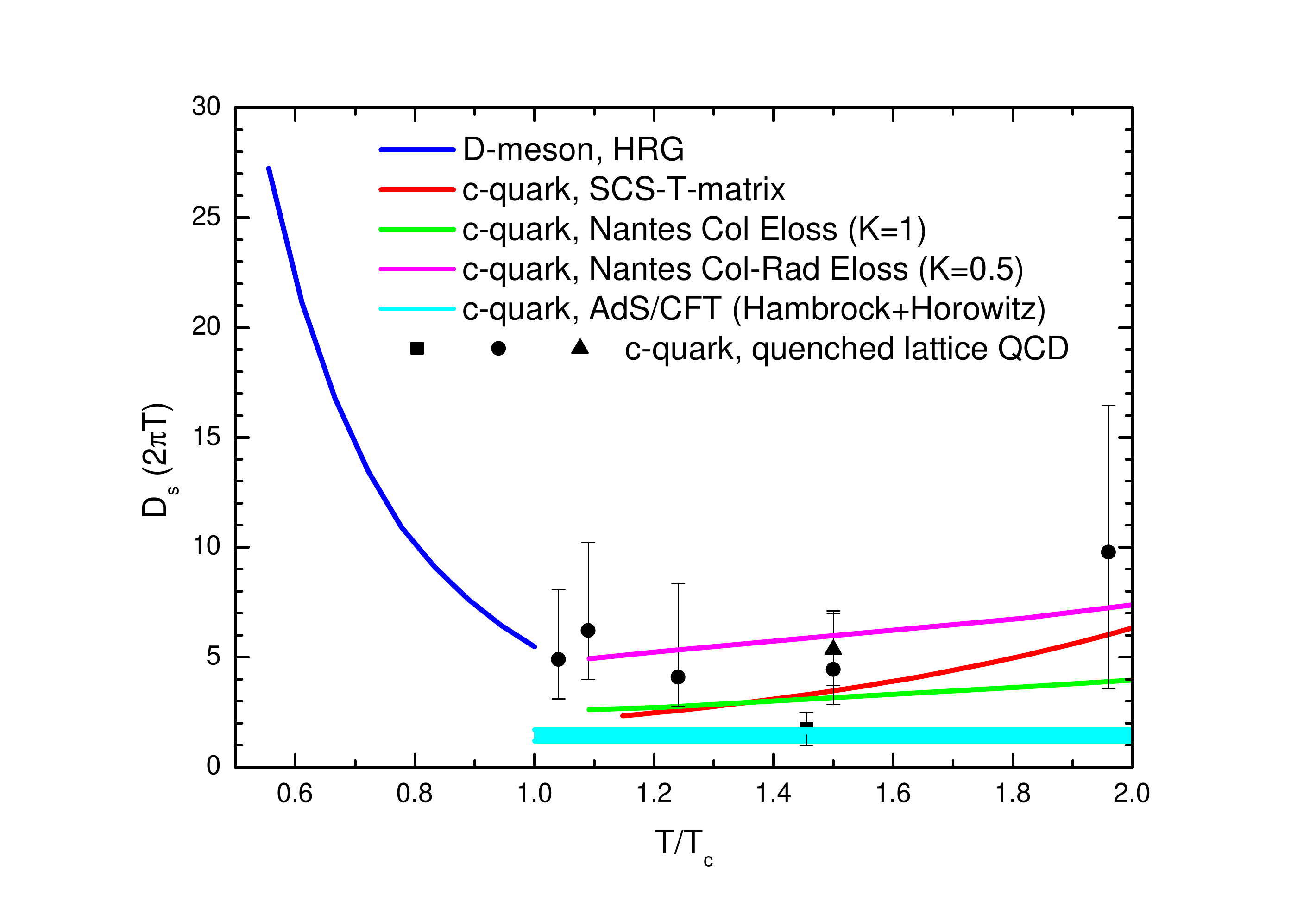}
\end{center}
\vspace{-0.8cm}
\caption{Spatial charm diffusion coefficient in the QGP ($T>T_c$) 
and hadronic matter ($T<T_c$) .
Updated results from this conference (Nantes~\cite{Gossiaux:2018} and 
AdS/CFT~\cite{Hambrock:2018sim}) are compared to recent $T$-matrix results 
(``Strong Coupling Scenario")~\cite{Liu:2017qah}, lQCD 
``data"~\cite{Ding:2011hr,Banerjee:2011ra,Francis:2015daa} and a hadronic 
calculation~\cite{He:2011yi}. Figure courtesy of M.~He.}
\label{fig_Ds}
\end{figure}

The impact of the space-time evolution has been discussed in Ref.~\cite{Gossiaux:2018};
even within bulk evolutin models (here EPOS-2 vs. EPOS-3) that 
give a good description of light-hadron observables, significant 
uncertainty remains in the extraction of the HQ diffusion 
coefficient. For the concrete case at hand, the EPOS-3 framework
tends to favor an increase of $\sim$30\%  of the ${\cal D}_s$ values in the Nantes 
energy-loss model (pQCD with running coupling and a rescaled Debye mass), bringing 
it, in fact, closer to values extracted by other groups, cf.~Fig.~\ref{fig_Ds}. 
This increase is at least in part caused by 
viscosity effects in EPOS-3 which lead to a slower cooling than
the ideal-hydro evolution used in EPOS-2~\cite{Gossiaux:2018}.
Also emphasized was the importance of controlled hadronization 
models, which for low- and intermediate-momentum observables 
should comply with the equilibrium limit in a locally thermalized medium. 
After all, hadronization is an interaction that should satisfy 
energy conservation and the central-limit theorem.   

The developments reported at this meeting reiterate the need for 
broadly based, yet detailed model comparisons, as have been 
commenced in a first round in 
Refs.~\cite{Rapp:2018qla,Cao:2018ews}. 
They also corroborate that the (scaled) HF diffusion 
coefficient ($2\pi T{\cal D}_s$) in QCD matter reaches values 
below 5, with a putative minimum in the vicinity of the 
transition temperature, and increasing with temperature and 
3-momentun reflecting the fundamental scale dependence of QCD.  

\section{Hadronization}
\label{sec_hadro}
The effects of charm-quark diffusion and hadronization through coalescence processes 
with light quarks from the comoving thermal medium are not easily disentangled from 
$\raa$ and $v_2$ observables, especially when limited to a single particle species 
(say, $D$-mesons). Thus, measurements of additional charm-hadron species are highly 
valuable to sort out the contributions of different mechanisms (likewise for bottom). 
An important benchmark are the production ratios in elementary $pp$ 
collisions. To begin with, it turns out that, at the LHC, the 
non-strange $D$-meson ratios (such as $D^+/D^0$ or 
$D^{*,+}/D^0$) measured in $pp$~\cite{Acharya:2017jgo} are compatible with predictions 
from the statistical hadronization model (SHM; including feeddown)~\cite{Andronic:2003zv}. 
This also holds for the $D_s/D^0$ ratio once a strangeness suppresion factor of 
$\gamma_s$$\simeq$0.5 is introduced, consistent with the well-known strangeness
suppression in the light-hadron sector. In AA collisions, the light $D$-meson ratios
remain consistent with the SHM, at least at low $p_T$. In addition, as strangeness 
production approaches equilibrium, an enhancement of $D_s$ production through 
recombination with chemically equilibrated strange quarks has been 
predicted~\cite{Andronic:2003zv,Kuznetsova:2006bh,He:2012df}, which appears to be 
realized at both RHIC~\cite{Xie:2018} and the LHC~\cite{Grosa:2018,Acharya:2018hre}.       

The challenge comes with the recently reported $\Lambda_c$ production data. 
In particular, the STAR collaboration~\cite{Xie:2018} 
measured a $\Lambda_c/D^0$ ratio in 10-80\%
Au-Au($\sqrt{s}$=200\,GeV)  collisions that is close 
to or even above one at 
relatively low $p_T\simeq 3$\,GeV (which is comparable to the $\Lambda_c$ mass 
implying that it represents a good fraction of the inclusive yield). As a 
consequence, the extracted $\Lambda_c$ yields account for about half of the 
produced charm quarks. At the LHC, a first measurement of this ratio in 
Pb-Pb($\sqrt{s}$=5.02\,TeV) collisions is also near one~\cite{Acharya:2018ckj}, 
although at a significantly higher average $p_T$$\simeq$7\,GeV. Attempts to 
describe these data within the resonance recombination model (which 
conserves 4-momentum and obeys the equilibrium limit) implemented on a 
hydrodynamic hypersurface have thus far failed~\cite{He:2011qa}. 
Instantaneous coalescence models can get rather close to the 
data~\cite{Oh:2009zj,Scardina:2017ipo} by utilizing nontrivial wavefunction 
effects in the hadronization process, allowing the $\Lambda_c/D^0$ ratio to 
exceed the equilibrium value by a large factor. This would imply that HQ 
hadronization proceeds rather far from equilibrium, with individual 
wavefunction effects for each hadron. It would be important to understand 
how this scenario can be reconciled with the light-hadron sector.

\section{Quarkonium Transport}
\label{sec_onia}
The (chemical-) equilibrium limit discussed in the previous
section is a pivotal ingredient in description of quarkonium 
transport in URHICs~\cite{Nardi:2018}. This is highlighted in the 
pertinent rate equation for the time evolution of the number of quarkonia,
$N_\cQ$,
\begin{equation}
\frac{dN_{\cQ}}{d\tau} = -\Gamma_{\cQ} [N_\cQ - N_\cQ^{\rm eq}] \ ,  
\label{rate}
\end{equation}
where $\Gamma_{\cQ}$ denotes the inelastic reaction rate 
which drives the quarkonium abundance towards it 
(temperature-dependent) equilibrium limit 
$N_\cQ^{\rm eq}$ (as given, \eg, by the SHM). 

Since the individual HQ spectra are not necessarily thermalized 
throughout the fireball evolution in URHICs,  corrections to the 
equilibrium limit in the regeneration of quarkonia (second term 
in Eq.~(\ref{rate})) need to be investigated. Pertinent progress 
has been reported in Ref.~\cite{Yao:2018dap}
where coupled Boltzmann equations for the diffusion of 
individual anti-/bottom quarks and inelastic bottomonium 
reactions have been solved.
Detailed balance at fixed temperature has been verified leading 
to a near exponential approach to the equilibrium limit, 
${\cal R}(\tau)=(1-\exp[\tau/\tau_b])$~\cite{Du:2017qkv},
where the time scale is given by the thermal relaxation time,
$\tau_b$, of $b$-quarks. With bottomonium binding energies 
calculated from a vacuum Coulomb potential, a good description 
of the $\Upsilon(1S,2S)$ data of 
CMS~\cite{Khachatryan:2016xxp,Sirunyan:2018nsz} and 
STAR~\cite{Ye:2017vuw} can be achieved. A modest contribution 
from regeneration is found, similar to Ref.~\cite{Du:2017qkv}.

Bottomonium transport has also been studied within the AdS/CFT 
correspondence using a Coulomb potential and strong-coupling 
dissociation rates~\cite{Barnard:2017tld}. The latter turn out 
to over-suppress the $\Upsilon(1S)$ yield relative to CMS data. 
The agreement is much improved using smaller, perturbative-QCD 
(pQCD) rates, not inconsistent with a compilation of results 
from the Kent-State~\cite{Strickland:2011aa}, Tsinghua~\cite{Liu:2010ej} 
and TAMU~\cite{Du:2017qkv} groups 
shown in Ref.~\cite{Rapp:2017chc}.   

A rate-equation approach using bottomonium binding energies 
from an in-medium complex Cornell potential, with additional break-up 
from gluo-dissociation, implemented into an 
ideal-hydro evolution, has been presented in 
Ref.~\cite{Hoelck:2016tqf}. Finite formation times of the 
$\Upsilon$ states, leading to a reduced suppression in the 
early phases, are found to be significant to obtain a fair 
description of the CMS data, leaving most of the inclusive 
$\Upsilon(1S)$ suppression due to feeddown contributions.

The significance of formation time effects in the transport of quarkonia in 
the early phases of the fireball -- treated schematically in most existing 
approaches -- calls for a more detailed investigation. Explicit quantum 
treatments of quarkonium transport, as discussed in Ref.~\cite{Escobedo:2018}, 
can more accurately address these effects, as well as the evolution leading 
up to the formation of the various bound states in the thermalized medium.  
It will be interesting to see whether quantum approaches, at both the single
HQ and the quarkonium level, can be remapped into phenomenologically
successful rate equation frameworks, and how large corrections to the current 
extraction of transport coefficients are.

In a slight generalization of the original idea of using the medium modifications 
of the $c\bar c$ binding into $J/\psi$ as a probe of deconfinement, the ultimate 
goal remains to utilize the in-medium spectroscopy of quarkonium states, through
their production systematics in URHICs, as a probe of the fundamental forces
in QCD matter. The current combined phenomenology of quarkonium transport already 
puts significant constraints on the in-medium binding energies, $E_B^\cQ(T)$, and 
dissociation rates, $\Gamma_{\cQ}$,  leading to a hierarchy in the bound state 
melting as (using the standard criterion 
$E_B^{\cQ}(T_{\rm melt}) \simeq \Gamma_{\cQ}(T_{\rm melt})$)
\beq
T_{\rm melt}[\psi(2S)] <  T_0^{\rm SPS} \lesssim  T_{\rm melt}[\jpsi,\ups(2S)]
\lesssim T_0^{\rm RHIC} < T_{\rm melt}[\ups(1S)] \lesssim T_0^{\rm LHC} \ ;
\eeq
Here, $T_0^{\rm SPS}$$\simeq$\,240\,MeV, $T_0^{\rm RHIC}$$\simeq$\,350\,MeV
and $T_0^{\rm LHC}$$\simeq$\,550\,MeV are initial temperatures 
as estimated, \eg, from hydrodynamic simulations or electromagnetic radiation. 
Similar hierarchies are also found in lattice-QCD based 
extractions using non-relativistic QCD~\cite{Kim:2018yhk} 
or thermodynamic $T$-matrix approaches~\cite{Liu:2017qah}.   
Clearly, the complexity of in-medium quarkonia renders them unsuitable as a 
thermometer; rather, with independent URHIC temperature information, their 
in-medium properties can be scrutinized, with the ultimate goal of determining 
the underlying interaction in QCD matter.

\section{High-$p_T$ Suppression}
\label{sec_highpt}
At high transverse momenta, HF particles provide a unique window on the  
characteristics of parton energy loss in the QGP. First, the transition 
from a collisional into a radiatively dominated regime can be studied (probably
not possible in the light sector where the transition momentum is likely in the
thermalized part of the spectrum); here, the factor of $\sim$3 difference in 
charm- and bottom-quark masses provides an extra leverage to identify this
regime.  Second, interference effects in gluon radiation can be scrutinized 
through the path length ($L$) dependence as discussed at this meeting
in Ref.~\cite{Djordjevic:2018ita}: starting from an ansatz 
for the fractional energy loss, $\Delta E/E \sim \eta T^a L^b$ (with a 
$p_T$-dependent coefficient $\eta$), the observable 
$R_L \equiv (1-R_{XeXe})/(1-R_{PbPb}) \simeq (A_{Xe}/A_{Pb})^{b/3}$ 
has been proposed as direct measure of the power $b$. Pertinent model
calculations~\cite{Djordjevic:2018ita} find a near linear dependence
($b$=1) for all flavors for $p_T$$\simeq$20\,GeV; for light flavors, it quickly 
develops into a significant nonlinear dependence with increasing $p_T$, 
but more gradually for $c$ and $b$ quarks, and eventually recovering flavor 
independence at $p_T$$\simeq$100\,GeV with $b$$\simeq$1.4.
A very close-to-linear dependence for $b$ quarks in the $p_T$=10-40\,GeV
range has also been found in the model comparisons conducted in 
Ref.~\cite{Rapp:2018qla}, while for $c$ quarks interference effects
are somewhat more pronounced for $p_T$$>$10\,GeV.

Furthermore, formation time effects arising from the finite virtuality in 
the production of $c$ and $b$ quarks can be studied at high $p_T$,
where they are augmented by Lorentz-$\gamma$ factors (similar to the
case of quarkonia discussed above, where, however, the relevant timescale, 
given by the inverse binding energies, is (much) longer). This was discussed 
in Ref.~\cite{Cao:2017crw}; in essence, the larger mass of the $b$ quarks
allows them to go on shell more quickly than $c$ quarks, enabling an 
additional quenching of the bottom $p_T$-spectra in the 10-50\,GeV range,
not inconsistent with high-$p_T$ suppression data.

\section{Small Systems}
\label{sec_small}
Both heavy quarks and quarkonia can contribute to a better understanding
of the mechanisms driving the apparent collectivity observed in small collision
systems, \ie, $p$/$d$A (and even high-multiplicty $pp$) collisions.
In Ref.~\cite{Vogt:2018oje}, it was reiterated that the modest modifications 
observed in the $D$-meson $\raa$ in $p$Pb collisions~\cite{Abelev:2014hha} can 
be well described by baseline shadowing calculations, in fact better than in 
models where $c$-quark interactions in a QGP are evaluated, which tend to cause 
too much suppression toward higher $p_T$. However, the main problem is the large 
$v_2$, reaching up to 10\% for $D$-mesons around $p_T$$\simeq$3.5\,GeV in 
high-multiplicity $p$Pb collisions~\cite{Sirunyan:2018toe}, which cannot be
explained by ``conventional" initial-state effects. It has been argued that the 
anisotropic escape effect, which is rather effective in generating light-hadron $v_2$ 
in small systems, is less effective for charm quarks~\cite{Li:2018leh}, rendering
them a better probe of medium collectivity. However, also in these calculations,
using the AMPT transport model with 3\,mb cross sections, large $c$-quark
$v_2$ values cannot be obtained for $p$Pb collisions. 

\section{Conclusions}
\label{sec_concl}
The ``Hard Probes 2018" meeting has demonstrated again the versatility and uniqueness 
of heavy-flavor hadrons to analyze key properties of QCD matter as produced in URHICs. 
Advances in understanding the different components in the modelling of low-momentum 
$D$-meson observables continue to narrow down the spatial diffusion coefficient of 
heavy quarks, a fundamental transport coefficient of the medium. The current range 
of extracted values, ${\cal D}_s(2\pi T)$=2-4, translates into a large scattering rates 
which imply that the medium's long-wavelength excitations do not support light-parton 
quasiparticles when approaching $T_{\rm pc}$ from above. This also offers a natural 
explanation why the hadronization region plays an important role in HF transport, and
goes hand-in-hand with the observed changes in the hadrochemistry of HF hadrons in 
URHICs, relative to $pp$ collisions, most notably the enhancements of $D_s$ mesons 
and $\Lambda_c$ baryons. For the latter, a satisfactory theoretical description remains
challenging. Part of the resoultion of this puzzle might reside in their unexpectedly 
large production in $pp$. A reliable understanding of the HF hadrochemistry is also 
pivotal to transport descriptions of charmonium production via their (re-) generation, 
where chemical equilibrium represents the long-time limit; at finite times, (re-)
generation is sensitive to the degree of the HQ thermalization.  
Going forward, the exploitation of these intricate connections promises for a tightly 
constrained framework of HF transport and hadronization, \ie, kinetics and chemistry. 
The large HQ mass serves as a control parameter that can be utilized to check 
approximations and tested via comparisons of charm and bottom observables. It also provides 
a scale, $p_{\rm trans}\sim m_Q$, where the non-pertubatively dominated low-momentum diffusion
physics transits into a perturbative regime dominated by energy loss via gluon radiation (with a 
possibly large coefficient). In the high-$p_T$ regime, quark mass dependencies seem to survive 
into the 10's of GeV regime. Finally, modifications of HF particles in small systems remain 
intriguing. The lack of a noticable $D$-meson suppression observed at high $p_T$, together 
with their large $v_2$ signal, pose a challenge for explanations based on final-state effects.  
Clearly, much remains to be learned about what HF observables tell us about the QGP
and its transition into hadrons.  
\\  

{\bf Acknowledgment}\\
I thank P.B.~Gossiaux and M.~He for interesting discussions.
This work has been supported by the U.S.~National Science Foundation under
grant PHY-1614484.

\end{document}